\documentclass[twocolumn,showpacs,amsmath,amssymb,prl]{revtex4-1}
\usepackage{graphicx} 
\usepackage{epsfig}
\usepackage{dcolumn} 
\usepackage{color}
\usepackage{changes}
\newcommand{\beq}{\begin{equation}}
\newcommand{\eeq}{\end{equation}}
\newcommand{\beqa}{\begin{eqnarray}}
\newcommand{\eeqa}{\end{eqnarray}}

\newcommand{\kvec}{{\bf k}}

\newcommand{\qvec}{{\bf q}}

\newcommand{\OO}{{\overline \Omega}}

\begin{document}
\title{
Pseudogap and (an)isotropic scattering in the fluctuating charge-density wave phase of cuprates}
\author{S. Caprara$^{1,2}$, C. Di Castro$^{1,2}$, M. Grilli$^{1,2}$, and G. Seibold$^3$}
\affiliation{$^1$ Dipartimento di Fisica, Universit\`a di 
Roma ``La Sapienza'', P.$^{le}$ Aldo Moro 5, 00185 Roma, Italy \\
$^2$ ISC-CNR and Consorzio Nazionale Interuniversitario per le Scienze Fisiche della 
Materia, Unit\`a di Roma ``Sapienza''\\
$^3$ Institut f\"ur Physik, BTU Cottbus-Senftenberg - PBox 101344, D-03013 Cottbus, Germany}

\begin{abstract}
We present a general scenario for high-tem\-pe\-ra\-ture superconducting cuprates, based on the presence 
of dynamical charge density waves (CDWs) and to the occurrence of a CDW quantum critical point, which occurs, e.g., 
at doping $p\approx 0.16$ in YBa$_2$Cu$_3$O$_{6+\delta}$ (YBCO). In this framework, even the pseudogap 
temperature $T^*$ is interpreted in terms of a reduction of the density of states due to incipient CDW and, at lower 
temperature to the possible formation of incoherent superconducting pairs. The dynamically fluctuating character of 
CDW accounts for the different temperatures at which the pseudogap ($T^*$), the CDW onset revealed by X-ray scattering 
($T_{ons}(p)$), and the static three-dimensional CDW ordering appear. We also investigate the anisotropic character 
of the CDW-mediated scattering. We find that this is strongly anisotropic only close to the CDW quantum critical point 
(QCP) at low temperature and very low energy. It rapidly becomes nearly isotropic and marginal-Fermi-liquid-like away 
from the CDW QCP and at finite (even rather small) energies. This may reconcile the interpretation of Hall measurements 
in terms of anisotropic CDW scattering \cite{arxiv} with recent photoemission experiments \cite{bok}.
\end{abstract}
 \date{\today}
\maketitle

\section{Introduction}
There are essentially two opposite points of view on the basic physical nature of the electronic state in
high-temperature superconducting cuprates. Since the early times the idea was put forward (mostly by P. W. Anderson, 
\cite{anderson1}) that these systems are strongly correlated doped Mott insulators, where the large electron-electron 
repulsion and the consequent short-range antiferromagnetic (AF) correlations, inside the low-dimensional layered 
structure of cuprates render these systems intrinsically different from standard metals ruled by the Landau 
Fermi-liquid (FL) paradigm.
The occurrence of this non-FL phase may imply a drastic rearrangement of the fermionic states: while far from the 
Mott state a FL is present with a large Fermi surface containing $n_h=1+p$ holes ($p$ is the doping) per unit cell 
in the CuO$_2$ planes, approaching the Mott state the metallic character is given by just $p$ carriers residing in four 
hole pockets in the so-called nodal regions $(\pi/2a)(\pm 1, \pm 1)$ of the Brillouin zone 
\cite{wen_lee,benfatto,YRZ,review_sachdev}. This transition is marked by a gradual
loss of spectral weight occurring below a $T^*(p)$ temperature, which vanishes at a critical doping $p^*\approx 0.19$. 
This is the famous pseudogap (PG) temperature, which in this context is clearly related to the ``Mottness'' of the 
metallic state.

The opposite point of view is that in two dimensions strong correlations and the proximity to a
doped Mott insulator are not enough to spoil the Landau FL and the anomalous behavior of metallic cuprates 
should be ascribed to the proximity to some form of instability ending at zero temperature into a QCP. In this case, 
the incipient order, which at low or zero temperature has an intrinsic quantum (and therefore dynamic) character 
produces strong long-ran\-ged and long-lived fluctuations. In turn these mixed quantum-thermal fluctuations
mediate strong scattering between the quasiparticles, spoiling the FL character of (some of) the quasiparticles 
and possibly mediating a strong superconducting pairing. In this ``quantum criticality'' scenario a crucial role 
is obviously played by the type of order that the system would like to realize, were it not for low dimensionality 
and disorder, which favor the competing presence of superconducting long-range order. Although many proposal have 
been put forward, the old evidences of CDWs \cite{reviewQCP1,reviewQCP2,kivelson_review} have been strongly revived 
by the recent ubiquitous observations of CDW  with NMR \cite{julien3}, STM \cite{yazdani,davis}, sound velocity
\cite{sound}, transport \cite{taillefer,sebastian,badoux}, and, most clearly, resonant X-ray scattering
\cite{ghiringhelli,chang2,cominscience2014,blancocanosa,croft,torchinsky,hucker3}. In XRS experiments CDW clearly 
emerge with an ordering vector $\qvec_c$ along the Cu-O bonds, i.e., along the (1,0) or (0,1) directions of the 
CuO$_2$ square lattice.

In this CDW framework, the CDW $\qvec_c$ precisely joins the portions of the Fermi surface where the PG first occurs. 
This naturally leads to the idea that CDW themselves can be responsible for the loss of the fermion spectral weight 
in the so-called antinodal regions of the Brillouin zone. In particular, it was recently shown \cite{arxiv} that the 
CDW quantum critical theory provides a coherent scenario rationalizing several issues, like the  
mechanism for CDW formation and its relation to the normal and superconducting state of cuprates, the mechanism  
fixing the direction of the CDW modulating wavevector $\qvec_c$ along the Cu-O bonds, the relation among the different 
CDW onset curves and corresponding QCPs, the relation between CDWs and PG, the role of CDWs in determining
the rapid change of the Hall number seen in experiments, the mechanisms leading to a dome-shaped CDW critical line,
delimited by two QCPs, at $p'_c\approx 0.08$ and $p_c\approx 0.16$. 

Starting from that analysis, whose main results are briefly recalled, we address here yet another relevant issue, 
namely: How isotropic or anisotropic is the nearly singular scattering mediated by CDW fluctuations and how this 
(an)isotropy mirrors in transport and spectroscopic experiments?

\section{Dynamical CDW }
\subsection{Fermi-liquid theory and the direction of the CDW wavevector}\label{model}
\begin{figure}[htb]
\includegraphics[width=9cm,clip=true]{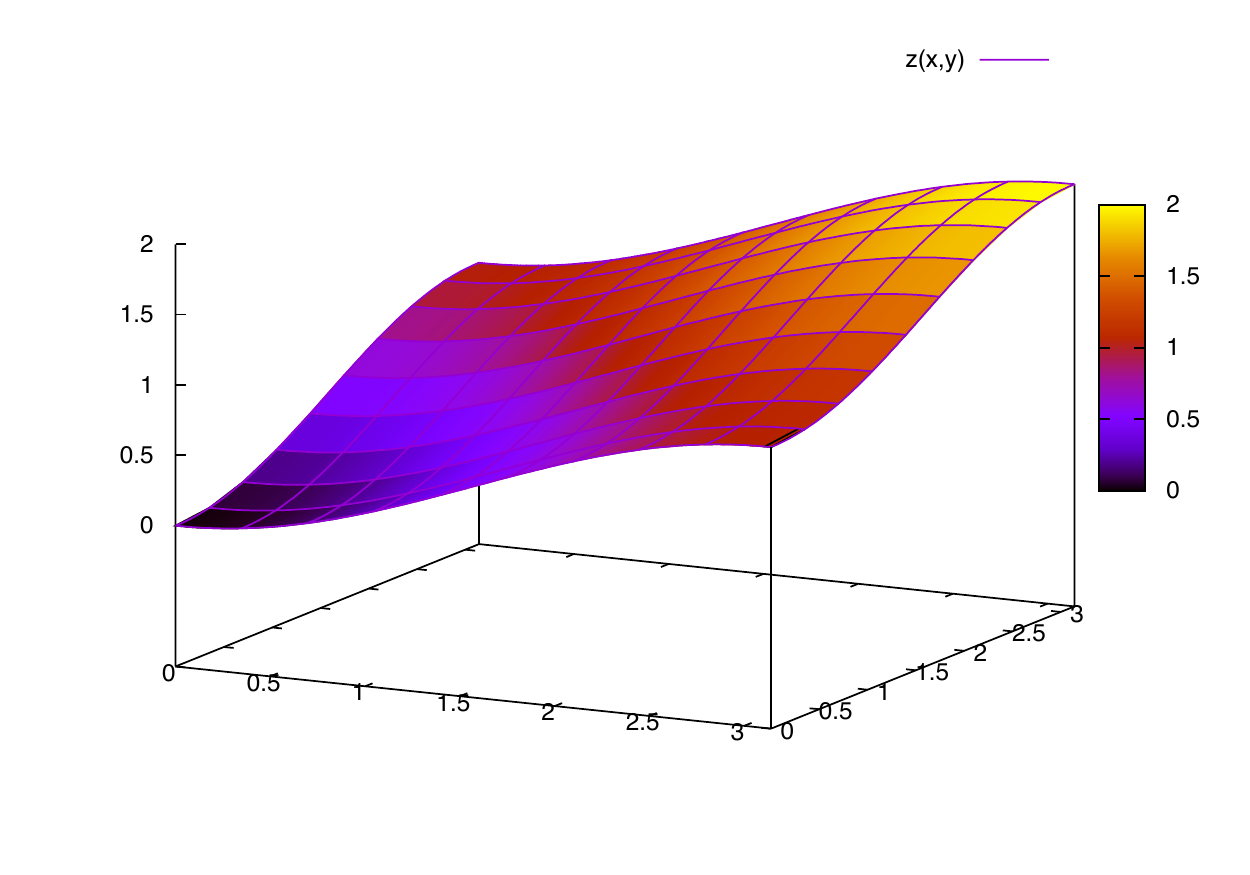}
\caption{Momentum dependence of $U(\qvec)$ for a $U=\infty$ single-band Hubbard model with nearest-neighbor hopping
$t=1$ and next-to-nearest-neighbor hopping $t'=-0.3t$, treated with slave bosons.}
\label{uq}
\end{figure}

Recent theories find that the CDW instability may be driven by retarded nearly critical 
spin fluctuations. However, in these approaches the CDW instability occurs along the  (1,1) direction 
\cite{sachdev,efetov}, in contrast with experiments, or requires a preliminary nematic deformation of the 
Fermi surface \cite{efetov2}. A recent functional renormalization group approach keeping into account the 
more detailed three orbital structure of the CuO$_2$ unit cell finds instead the right instability direction 
\cite{kontani} as it also happens when hole pockets are assumed as a prerequisite from a nearly ordered AF state
\cite{atkinson}. While all these approaches strongly rely on retarded collective spin fluctuations, 
a CDW mechanism was proposed long ago, which is based instead on the rather common tendency of strongly
correlated systems to become unstable under phase separation when even mild attractive forces mediated
by short-range AF coupling and/or phonons are present \cite{CGK,GRCDK}. Of course, this tendency is hindered 
by the long-range Coulomb repulsion between the charged quasiparticles and the system then choses
a compromise with short-range charge inhomogeneity. In this case CDW naturally arise \cite{CDG,becca,becca1998}.
More specifically, within a standard Random Phase Approximation closely mimicking the strong-correlation 
approaches (slave-boson or Gutzwiller), one can see that the instability occurs when the denominator of the 
density-density response function vanishes, as fixed by the condition $1+V(\qvec)\Pi(\qvec,\omega_n)=0$,
at zero Matsubara frequency, $\omega_n=0$, and $\qvec=\qvec_c$  \cite{CDG,reviewQCP1,reviewQCP2,andergassen}. 
Here, the residual interaction among the
quasiparticles $V(\qvec)= U(\qvec)-\lambda+V_c(\qvec)$ arises from three distinct contributions \cite{epj2000,arxiv}:
$U(\qvec)$ is a short-range residual repulsion resulting from the bare large repulsion of a one-band Hubbard model, 
$\lambda$ is a weakly momentum dependent short-range attraction promoting charge segregation (it may be due to a 
local phonon \cite{CDG}, to the instantaneous magnetic interaction present in doped antiferromagnets, or to 
both mechanisms), and $V_c(\qvec)$ is the long-range part of the Coulomb interaction. The screening processes 
are customarily described by the Lindhard polarization bubble $\Pi(\qvec,\omega_n)$ for quasiparticles having 
a renormalized band structure fitting the dispersion obtained from angle-resolved photoemission spectroscopy 
(ARPES) experiments. Two major points should be appreciated at this stage. First of all, the driving force for the 
CDW formation is the natural tendency to phase separation, which is triggered by {\it non critical} interactions, 
phonons and/or short-range AF. Therefore the proximity to an AF QCP is immaterial here and the Mottness only acts 
to weaken the metallic character of the FL quasiparticles, making them more easily prone to phase separation. 
Secondly, the short-range residual interaction between the FL quasiparticles has a momentum structure, which is
reported in Fig.\,\ref{uq} for a typical case.

This structure has nothing critical, but is the mere result of the screening processes bringing the strongly 
interacting electrons to dressed quasiparticles with moderate/weak residual interaction. Since the kinetic energy 
is typically larger along the (1,0) and (0,1) directions, quite naturally the residual repulsion $U(\qvec)$ is 
smaller along these direction. Therefore the instability condition is more easily realized along the Cu-O bonds. 
This mechanism therefore provides a robust tendency to establish the CDW order along these directions, in agreement 
with RIXS and XRD experiments.

\subsection{Pseudogap,  CDW threedimensional static order, and CDW onset temperature}
Expanding $V(\qvec)$ and $\Pi(\qvec,\omega_n)$ around $\qvec=\qvec_c$ and $\omega_n=0$ one obtains the 
standard quantum-critical charge-fluctuation propagator
\beq
D(\qvec,\omega_n)=\frac{1}{m_0+\nu(\qvec)  +|\omega_n| +\displaystyle{\frac{\omega_n^2}{\overline{\Omega}}}},
\label{fluctuator}
\eeq
where $m_0\propto 1+V(\qvec_c)\Pi(\qvec_c,0)$ is the mean-field {\it mass} of the fluctuations, and is proportional 
to the inverse of the square correlation length of CDW fluctuations, 
$\nu(\qvec)\approx \bar\nu |\qvec -\qvec_c|^2$ is the dispersion law of Landau-damped CDW fluctuations, 
$\bar\nu$ is an electronic energy scale (we work with dimensionless momenta, measured in inverse lattice spacings 
$1/a$), and $\overline{\Omega}$ is a frequency cutoff, setting the frequency above which CDW fluctuations become 
more propagating. The mean-field instability line $T^0_{CDW}(p)$ is characterized by a vanishing $m_0$, and corresponds
to the line in the temperature vs. doping diagram where $1+V(\qvec_c)\Pi(\qvec_c,0)=0$.
According to the scenario presented in Ref. \cite{arxiv}, $T^0_{CDW}(p)$ tracks the PG onset line $T^*(p)$. 
Therefore, in this scheme, $T^*$ is not related to any exotic realization of non-FL states, but simply occurs because 
CDW fluctuations start to deplete the states  around the antinodal region of the Fermi surface. This explains 
the identification of our theoretical mean-field line $T^0_{CDW}(p)$ with the experimental PG line $T^*(p)$, ending
into a missed QCP at $p^*\approx 0.19$. 

The fluctuation suppression of the mean-field critical line $T^0_{CDW}(p)$ is obtained by the self-consistent 
evaluation of the correction to the mean-field mass $m_0$, due to the fluctuator Eq. (\ref{fluctuator}). In this 
way, a renormalized mass $m(T,p)=m_0(T,p)+\delta m(T,p)$ is obtained, which vanishes
at the true CDW ordering temperature $T_{CDW}(p)$, which ends into a QCP at $p_c\approx 0.16$. It is worth noticing 
that in a strictly two-dimensional system, the renormalized ordering temperature for incommensurate CDW would 
vanish. However, in real layered systems, as cuprates, the planes are weakly coupled and, introducing a small energy 
scale $\nu_\perp$ related to the inter-plane coupling, a finite $T_{CDW}(p)$ can be obtained. This temperature, 
however, is so strongly reduced with respect to the mean-field line $T_{CDW}^0(p)$ that it occurs below the superconducting 
dome. Superconductivity therefore appears as the stabilizing phase against CDW long-range order. This explains why 
the experimental data corresponding to long-range CDWs are only detected for magnetic fields large enough to weaken 
the superconducting phase \cite{sound}. 

\begin{figure*}[htb]
\includegraphics[width=15cm,clip=true]{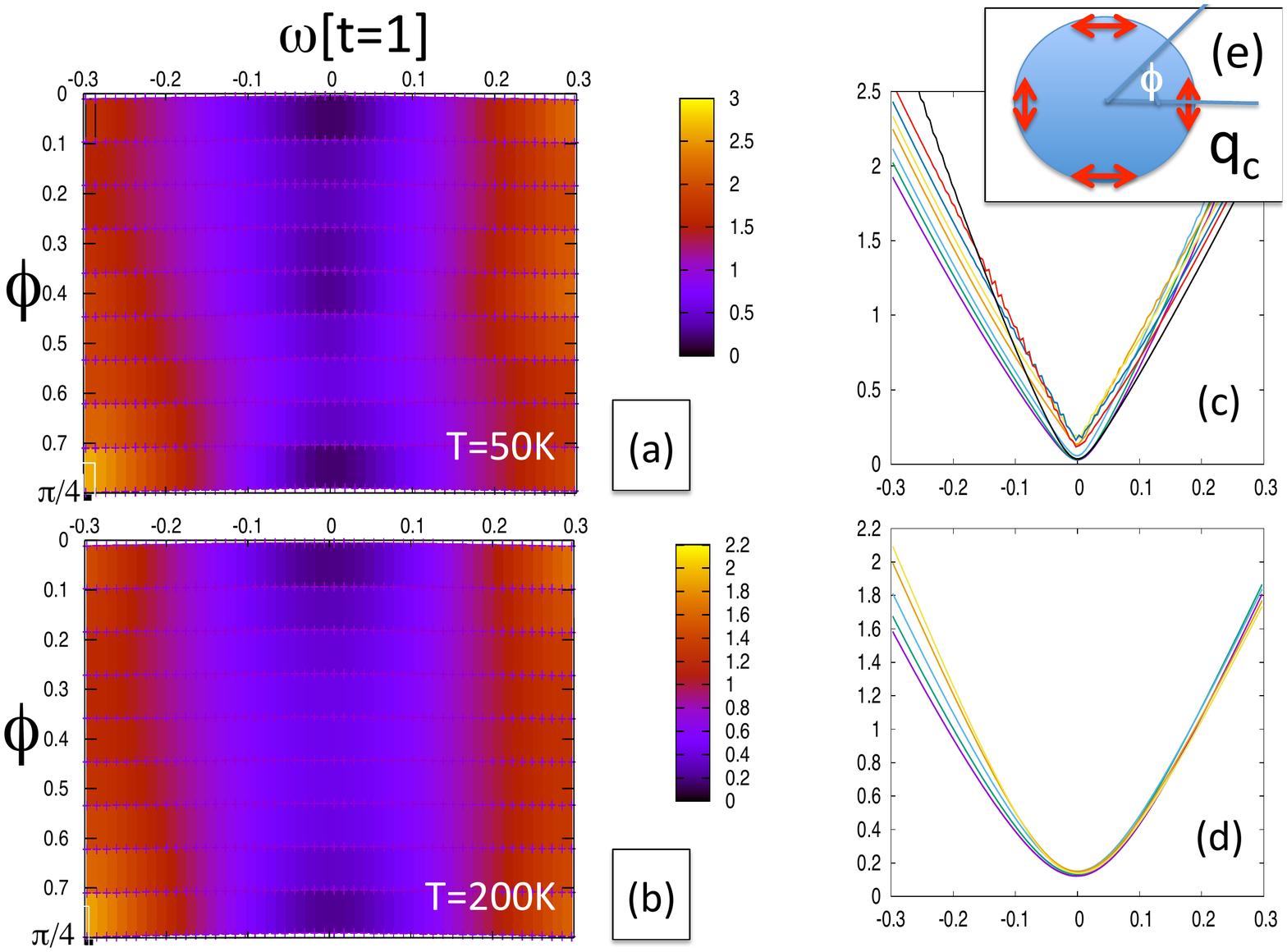}
\caption{Imaginary part of the electronic self-energy $\Sigma''\equiv \mathrm{Im}\Sigma(\phi,\omega)$ due to 
CDW fluctuation scattering at the lowest order in perturbation theory, with $T=50$\,K (a) and $T=200$\,K (b). 
Selected $\Sigma''(\omega)$ curves at different $\phi$, equally spaced in the interval $[0,\frac{\pi}{4}]$, at 
$T=50$\,K (c) and $T=200$\,K (d). (e) Sketch of the isotropic Fermi surface, of the CDW $\qvec_c$'s, and 
definition of the angle $\phi$.}
\label{fig2}
\end{figure*}

While at doping $p\gtrsim p_c$, the complex CDW order parameter fluctuates in modulus and phase, a different 
situation is met at low doping, where the physics is dominated by fluctuations of the phase of the CDW complex 
order parameter only \cite{arxiv}. Thus the critical line $T_{CDW}(p)$, that would be monotonically increasing
with decreasing doping, in the case ruled by modulus and phase fluctuations, becomes dome-shaped due to the
prominent role of phase fluctuations at low doping, and ends into yet another QCP at $p_c'\approx 0.08$, in agreement
with the experiments.

It must be noted that in experiments CDW appear in two ways: when fast probes like X-rays are used a dome-like 
onset temperature 
$T_{ons}(p) \approx 100-150$\,K is identified. On the other hand, static CDW are found both with fast and slow
probes at the lower $T_{CDW}(p)$. This raises the intriguing issue is why different probes identify different 
CDW onset temperatures. The key point is the dynamical character of the CDW fluctuations. A probe with very long 
characteristic timescale $\tau_{pr}$  (like, e.g., NMR or NQR) will only detect static order, otherwise the 
fluctuating CDWs average to zero during the probing time. This is why these probes identify a true phase-transition 
line $m(T_{CDW},p)=0$ at high magnetic field (of course, if in real systems pinning intervenes to create locally a 
static order, this can be detected by static probes even at larger temperatures and low magnetic fields \cite{julien3}). 
On the other hand, a fast probe with a short probing time $\tau_{pr}$ takes a fast snapshot of the fluctuating 
system and finds a higher ``transition'' temperature when the CDW order is still dynamical, as long as the 
CDW characteristic timescale $\tau_{CDW}\propto \xi^2\propto m^{-1}$ is longer than $\tau_{pr}$. We thus identify 
the dynamical onset line as the line where $m$ reaches its minimal dynamical value 
$m\approx\omega_{pr}=\tau_{pr}^{-1}$ \cite{arxiv}.

On the one hand, this qualitatively solves  the experimental puzzle accounting for different CDW onset lines. 
However, the value $\omega_{probe} \sim 50-100$\,K needed to fit the CDW onset line detected with RIXS \cite{arxiv}
corresponds to timescales of order of $0.1$\,ps that are still too long in comparison with the fast (of order of a 
few fs) characteristic times of RIXS. We notice that this discrepancy could be reconciled if the onset temperature 
found by these fast probes were coinciding with $T^0_{CDW}(p)$ (i.e. all CDW fluctuations are slower than the 
RIXS snapshot) which is $50-100$\,K higher. We therefore suggest that a higher sensitivity of these experiments 
could shift at substantially higher temperatures the detection of the CDW peak, much closer to the $T^*$ line.

\section{(An)isotropy of the quasiparticle scattering}
The proximity to a CDW QCP provides an anisotropic scattering mechanism between the FL quasiparticles quite similar 
to the one due to spin fluctuations near an AF QCP \cite{abanov}. In Ref. \cite{arxiv} it was shown that this 
anisotropic scattering may account for the rapid changes observed in the Hall constant 
$R_H$ of YBCO at low temperature, when the doping is increased from $p\gtrsim 0.16$ to $p=0.19$. The rapid growth 
of $R_H$ with increasing doping is naturally interpreted in terms of a large Fermi surface corresponding to a hole 
density $1+p$ reconstructing to hole pockets enclosing $p$ states only. However, any reconstruction of a Fermi 
surface must occur via a change in 
the electronic state. While the possibility can be considered that the pockets are formed in an exotic 
Mottness-driven non-FL state, it is also natural to associate this new state to the CDW QCP occurring at $p=0.16$, 
when the ordered CDW state takes place under strong magnetic field. In this latter framework, increasingly slower 
and more extended CDW correlations are present when the doping is reduced from $p=0.19$ to $p =0.16$. This implies 
that at least in the initial steps of this underdoping the reduction of $R_H$ should be interpreted as due to 
the increasingly stronger and anisotropic CDW mediated interactions. This allowed for the successful fitting of 
$R_H$ in terms of such increasingly anisotropic interactions. Of course, other effects neglected in our analysis,
like the interplay with CDW-mediated pairing correlations, could gradually add their contribution in this crossover 
region, possibly promoting the formation of Fermi arcs. However, the experiments cannot exclude that the Fermi surface
stays large at zero temperature.

On the other hand, ARPES experiments were recently carried out \cite{bok} on Bi$_2$Sr$_2$CaCu$_2$O$_{8+\delta}$ 
(Bi2212) samples, finding indications of an isotropic electronic self-energy compatible with a marginal-FL state 
\cite{marginal}. At first sight this result seems more naturally explained by a microscopic model based on a 
$\qvec_c=0$ instability like the one due to time-reversal symmetry breaking loop-current fluctuations \cite{varma} 
and seems at odds with the anisotropic singular scattering mediated by CDW fluctuations. This section is precisely 
devoted to a possible solution of this puzzle.

\begin{figure}[htb]
\includegraphics[width=8cm,clip=true]{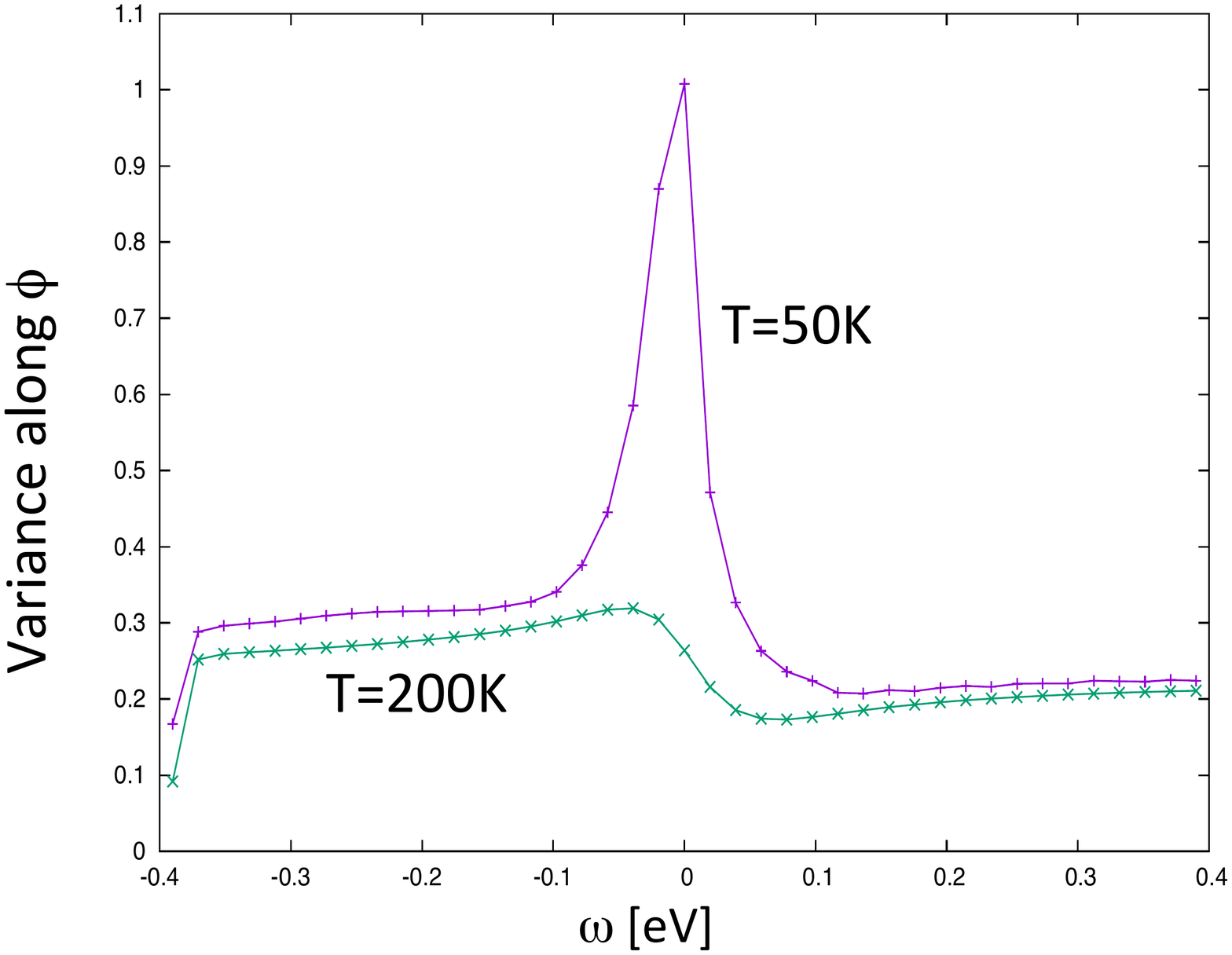}
\caption{Variance $\sigma(\omega)$ of $\Sigma''(\phi,\omega)$ averaged over $\phi$ around the Fermi surface at low 
$T=50$\,K (purple curve) and intermediate $T=200$\,K (green curve).}
\label{fig3}
\end{figure}

First of all, to get rid of ``trivial'' anisotropy due to the band structure, we purposely consider a simple 
isotropic pa\-ra\-bo\-lic electronic band $\epsilon_\kvec=-E_0+t k^2$, where $E_0=1$ eV and $t=0.25$ eV (again $k$ 
is dimensionless). This gives rise to a large circular Fermi surface of quasiparticles with mass $m_{qp}=1/2t$. 
Then, taking the typical parameters of the CDW fluctuator (\ref{fluctuator}) the lowest-order self-energy correction 
is calculated to be compared with that obtained from ARPES 
\beqa
&~& \mathrm{Im}\Sigma(\kvec,\omega) =  \\
&~& g^2\int \frac{d \qvec}{(2\pi)^2} \frac{ (\omega -\epsilon_{\kvec-\qvec}) \left[ b(\epsilon_{\kvec -\qvec}) 
+ f(\epsilon_{\kvec-\qvec}-\omega)) \right]}{ \left[ \nu(\qvec) - (\omega -\epsilon_{\kvec-\qvec})^2/\OO \right]^2 
+  (\omega -\epsilon_{\kvec-\qvec})^2 }
\nonumber
\eeqa
where $g$ is the quasiparticle-CDW coupling. Here, $\qvec_c$ is arbitrarily chosen to connect different parts of 
the perfectly isotropic Fermi surface, in order to introduce the effects of the anisotropic CDW-mediated interaction 
[see Fig. \ref{fig2}(e)]. To emphasize this effect, we select the mass of the CDW corresponding to $p=0.16$, as 
determined from the analysis of Ref. \cite{arxiv}, which vanishes at $T=0$. Fig. \ref{fig2} reports 
$\mathrm{Im} \Sigma=\Sigma''$ as a function of energy for $|\kvec| =k_F$ and various angles $\phi$ along the Fermi 
surface. One can see that at the angles $\phi_{hot}$ corresponding to the hot spots (where points of the Fermi surface are 
connected by $\qvec_c$) $\Sigma''(\phi_{hot}, \omega)\sim \sqrt{\omega}$, according to previous perturbative analyses 
\cite{abanov,sulpizi}. This is clearly visible in the panels (c) (green and light-blue curves), where the mass in the 
CDW  fluctuator is small because at low temperature ($T=50$\,K) the system is close to the CDW QCP. On the other hand, 
as soon as one moves away from the hot spots, the behavior of $\Sigma''$ rapidly recovers the quadratic shape typical
of a FL. When the temperature grows to $T=200$\,K, the mass of the CDW fluctuations increases (i.e., the correlation
lenght $\xi_{CDW} \sim 1/\sqrt{m}$ decreases) and $\Sigma''(\phi,\omega)$ becomes much more isotropic. This can 
be quantified by calculating $\sigma(\omega)$, defined as the variance of $\Sigma''( \omega)$ when this is averaged 
over $\phi$ around the Fermi surface. Fig. \ref{fig3} shows that  $\sigma(\omega)$ is usually small (of order 
20-30 percent) both at low and high temperature. 

The only large variations in $\phi$ occur at low energies (smal\-ler than $50$\,meV) and low temperature, because in 
this case a strong anisotropy is present due to the nearly singular scattering at the hot spots. These results 
clearly solve the (an)\-iso\-tro\-py puzzle because they show that the CDW scattering is only quasi-singular and produces 
non-FL behavior in narrow regions around the hot spots and at low energies. This accounts for the strong effects 
observed in transport experiments, like in the Hall effect \cite{badoux,arxiv}. On the other hand, at finite 
frequencies the scattering is rather strong but nearly isotropic over the whole Fermi surface. We also notice that 
$\Sigma''$ becomes rather isotropically linear in frequency, mimicking well a marginal FL behavior.
Of course in the isotropic band we purposely adopted, without an upper limit to the energy, no ultraviolet 
cutoff intervenes to stop this marginal-FL behavior at positive frequencies.

\section{Conclusions}
Our work shows that an internally coherent scenario is possible in which $T^*$ marks the initial appearance of 
CDW fluctuations. Moreover, as it was shown long ago \cite{perali}, CDW may mediate d-wave pairing and therefore 
an additional mechanism of PG formation due to pairing may intervene when the fluctuating CDW glue becomes strong 
enough. Of course, according to the Mottness supporters, the possibility is still open, that CDW are just an 
``epiphenomenon'' occurring on top of the more fundamental physics ruled by strong correlations spoiling the FL 
and that the origin of $T^*$ has nothing to do with CDW. The point of view of our work is instead fully within 
the ``quantum criticality'' scheme showing that the PG occurring below $T^*$ and all the CDW phenomenology (like, 
e.g., the Fermi surface reconstructions revealed by transport in strong magnetic field) may all stem from the 
occurrence of CDW around optimal doping $p\approx 0.16$ (in YBCO) and are not tightly related to the Mott physics 
and to the disappearance of long-range antiferromagnetism occurring at much lower doping ($p=0-0.02$).

\end{document}